\documentclass[%
reprint,
 amsmath,amssymb,
 aps,
]{revtex4-1}

\usepackage{graphicx}
\usepackage{dcolumn}
\usepackage{bm}
\usepackage{dsfont}
\usepackage{amsthm}
\usepackage[dvipsnames]{xcolor}
\usepackage{chngcntr}
\usepackage{apptools}
\AtAppendix{\counterwithin{lemma}{section}}
\usepackage{longtable} 

\usepackage[capitalise]{cleveref}
\crefname{figure}{Fig.}{Fig.}

\newtheorem{theorem}{Theorem}

\newtheorem{lemma}{Lemma}
\theoremstyle{definition}
\newtheorem{definition}{Definition}
\theoremstyle{theorem}

\theoremstyle{definition}

\newcommand{\eps}{\epsilon}

\begin{document}

\title{Classical Simulation of Noncontextual Pauli Hamiltonians}

\author{William M. Kirby}
\affiliation{Department of Physics and Astronomy, Tufts University, Medford, MA 02155
}%
\author{Peter J. Love}
 \altaffiliation[Also at ]{Brookhaven National Laboratory}
\affiliation{Department of Physics and Astronomy, Tufts University, Medford, MA 02155
}%

\begin{abstract}
Noncontextual Pauli Hamiltonians decompose into sets of Pauli terms to which joint values may be assigned without contradiction.
We construct a quasi-quantized model for noncontextual Pauli Hamiltonians.
Using this model, we give an algorithm to classically simulate noncontextual VQE.
We also use the model to show that the \textsc{noncontextual Hamiltonian problem} is NP-complete.
Finally, we explore the applicability of our quasi-quantized model as an approximate simulation tool for contextual Hamiltonians.
These results support the notion of noncontextuality as classicality in near-term quantum algorithms.
\end{abstract}

\pacs{Valid PACS appear here}
\maketitle

\section{Introduction}
\label{intro}

Simulation of quantum systems using the \emph{variational quantum eigensolver} (VQE) is a promising application for noisy intermediate-scale quantum (NISQ) computers \cite{preskill18a,peruzzo14a}.
VQE is advantageous in the NISQ era because the necessary circuit depths are small compared with other quantum simulation methods, such as phase estimation \cite{peruzzo14a}.
Small-scale VQE experiments have been performed in a variety of qubit architectures, to simulate systems from molecular to high-energy physics \cite{DuH2NMR,lanyon2010towards,peruzzo14a,wang2015quantum,omalley16a,santagati18a,shen2017quantum,paesani2017,kandala17a,hempel18a,dumitrescu18a,colless18a,nam19a,kokail19a,kandala19a}.

In this paper, we give a classical simulation technique for \emph{noncontextual Hamiltonians}, defined in our previous work \cite{kirby19a}.
Contextuality is an indicator of nonclassicality \cite{bell64a,bell66a,kochen67a} that is related to negativity of quasi-probability representations \cite{spekkens08,ferrie08a,ferrie09a,ferrie11a,veitch12a,raussendorf17a} and resources for quantum computation \cite{raussendorf13a,howard14a,karanjai18a,raussendorf19c,schmid18a,duarte18a,okay18a,frembs18a,arvidsson-shukur19a}.
Recent work has led to a variety of approaches to characterizing and quantifying contextuality \cite{abramsky11a,ramanathan12a,cabello14a,ramanathan14a,grudka14a,cabello15a,abramsky17a,de_silva17a,amaral17a,xu18a,mansfield18a,kirby19a,raussendorf19b,schmid19a}.

In the present work, we show that when a Hamiltonian is noncontextual according to the criterion of \cite{kirby19a}, we may construct a \emph{quasi-quantized} classical model of the associated VQE instance.
Quasi-quantized models are descriptions of classical subtheories of quantum mechanics, such as the stabilizer subtheory for odd-dimensional qudits, and Gaussian subtheory for continuous systems \cite{spekkens16a,spekkens07a,vanenk07a,spekkens08,bartlett12b}.

Our quasi-quantized model is a hidden-variable theory for prepare-and-measure scenarios where all states are allowed but the measurements must correspond to a noncontextual set of Pauli observables.
This contrasts with \cite{raussendorf19b}, in which it is shown that for noncontextual states (in the sense of being positively representable over a noncontextual phase space), any set of Pauli measurements permits classical description.

The problem of approximating the ground state energy of a classical or quantum  Hamiltonian can be NP-complete or QMA-complete, respectively~\cite{barahona82a,kitaev02,wocjan03a,kempe03a,bravyi05a,kempe06a,oliviera08a,biamonte2008realizable,aharonov11a,schuch11a,yan12a,aharonov15a}. Our quasi-quantized model allows us to show that the \textsc{noncontextual Hamiltonian problem} (see \cref{complexity_intro}) is only NP-complete, rather than QMA-complete~\cite{kitaev02}. In other words, the fact that we can describe the VQE procedure in terms of nonnegative joint probabilities places noncontextual VQE unambiguously in the realm of the classical, and this result extends to the computational complexity of the problem. 

In \cref{model}, we show how to construct a quasi-quantized model for any noncontextual Hamiltonian.
The states of the quasi-quantized model are probability distributions that correspond to quantum states.
We prove that these probability distributions reproduce expectation values for the Hamiltonian terms, including the expectation values corresponding to Hamiltonian eigenstates.
In \cref{simulation}, we use our model to construct a classical simulation algorithm for VQE.
We also show that the \textsc{noncontextual Hamiltonian problem} is NP-complete.
In \cref{approx_sect}, we show how to approximate contextual Hamiltonians by noncontextual Hamiltonians.
In some cases we reach chemical accuracy using this approximation, and in most cases we outperform existing experimental results.
In \cref{discussion}, we summarize and discuss our results.

\subsection{Variational Quantum Eigensolvers}
\label{vqe_intro}

In VQE, we wish to minimize the energy expectation value of a Hamiltonian $H$:
\begin{equation}
    \label{vqehamiltonian}
    \langle H\rangle=\sum_{P\in\mathcal{S}}h_P\langle P\rangle,
\end{equation}
where the $h_P$ are real coefficients, and $\mathcal{S}$ is the set of Pauli operators in the support of the Hamiltonian (see \cite{mcardle18a} for a review of VQE).
The energy expectation value~\eqref{vqehamiltonian} is estimated by preparing a physical ansatz on a quantum device, and evaluating the expectation value of each Pauli term $P\in\mathcal{S}$ separately. The weighted sum \eqref{vqehamiltonian} is then treated as an objective function for a classical optimization of the ansatz parameters~\cite{peruzzo14a}.

\subsection{Local Hamiltonian Problems}
\label{complexity_intro}

The $k$-\textsc{local Hamiltonian problem} is the decision problem of whether the ground state energy of a $k$-local Hamiltonian lies below some specified energy gap (with size at least $1/\text{poly}(n)$, for $n$ qubits), given the promise that the ground state does not lie within the gap \cite{kitaev02}.
The $k$-\textsc{local Hamiltonian problem} is QMA-complete, and has also been studied for various subsets of Hamiltonians (such as $k$-local commuting Hamiltonians, and Hamiltonians with specific interaction graphs) \cite{barahona82a,kitaev02,wocjan03a,kempe03a,bravyi05a,kempe06a,oliviera08a,biamonte2008realizable,aharonov11a,schuch11a,yan12a,aharonov15a}.
The complexity of the \textsc{$k$-local commuting Hamiltonian problem} is a long-standing open problem \cite{bravyi05a,aharonov11a,schuch11a,yan12a,aharonov15a}.

In this paper we focus on Pauli Hamiltonians. The complexity of the corresponding commuting problem is known: the \textsc{$k$-local commuting Pauli Hamiltonian problem} is in NP, since it is a special case of the \textsc{$k$-local commuting, factorizable qubit Hamiltonian problem}~\cite{bravyi05a}~\footnote{Note that this problem is distinct from that shown to be in P in \cite{yan12a}, in which the Hamiltonian is a linear combination of commuting Pauli operators \emph{with coefficients $\pm1$}}.
Since all commuting Pauli Hamiltonians are noncontextual, we will generalize this by proving that the \textsc{noncontextual Hamiltonian problem} is NP-complete.

\subsection{Noncontextual Hamiltonians}
\label{noncontextual_intro}

For $\mathcal{S}$ the set of Pauli operators in the Hamiltonian, let $\overline{\mathcal{S}}$ denote the \emph{closure under inference} of $\mathcal{S}$ \cite{kirby19a,raussendorf19b}.
This means that $\overline{\mathcal{S}}$ is the smallest set containing $\mathcal{S}$ as a subset, such that for every commuting pair $A,B\in\overline{\mathcal{S}}$, $AB\in\overline{\mathcal{S}}$ as well.
Since we could in principle measure $A$ and $B$ simultaneously together with their product $AB$, from any value assignment to $A,B$ we may \emph{infer} the assignment to $AB$: hence closure under inference.
We say that a value assignment is \emph{consistent} if it respects all such inference relations.
A value assignment to the observables in $\mathcal{S}$ (an \emph{ontic state}) extends to an assignment to $\overline{\mathcal{S}}$, and noncontextuality is defined by situations in which consistent assignments to $\overline{\mathcal{S}}$ exist \cite{kirby19a}.

We can determine whether $\mathcal{S}$ is contextual by the following criterion.
Let $\mathcal{Z}\subseteq\mathcal{S}$ be the set of operators in $\mathcal{S}$ that commute with all operators in $\mathcal{S}$
(we say that elements of $\mathcal{Z}$ commute universally in $\mathcal{S}$).
Let $\mathcal{T}\equiv\mathcal{S}\setminus\mathcal{Z}$; then \emph{$\mathcal{S}$ is noncontextual if and only if commutation is an equivalence relation on $\mathcal{T}$} \cite{kirby19a}.
In other words, if $\mathcal{S}$ is noncontextual, then $\mathcal{T}$ is a union of $N$ disjoint ``cliques" $C_1,C_2,...,C_N$, such that operators in different cliques anticommute, while operators in the same clique commute.

Let $C_i\equiv\{C_{ij}~|~j=1,2,...,|C_i|\}$.
A noncontextual Hamiltonian then has the form
\begin{equation}
    \label{hami}
    H=\sum_{i=1}^N\left(\sum_{j=1}^{|C_i|}h_{ij}C_{ij}\right)+\sum_{B\in\mathcal{Z}}h_BB,
\end{equation}
where $h_{ij}$ and $h_B$ are real coefficients.

We may rewrite this Hamiltonian in a useful way by defining $A_{ij}\equiv C_{ij}C_{i1}$.
$A_{ij}$ is itself a Pauli operator (up to a sign) and commutes universally in $\mathcal{S}$, since any operator in $C_i$ or in $\mathcal{Z}$ commutes with both $C_{ij}$ and $C_{i1}$, and any operator in one of the other cliques anticommutes with both $C_{ij}$ and $C_{i1}$.
Thus, we may rewrite \eqref{hami} as
\begin{equation}
    \label{ham}
    H=\sum_{i=1}^N\left(\sum_{j=1}^{|C_i|}h_{ij}A_{ij}\right)C_{i1}+\sum_{B\in\mathcal{Z}}h_BB.
\end{equation}
Note that in \eqref{ham}, the only operators appearing on the right-hand side that do not commute universally in $\mathcal{S}$ are the $C_{i1}$: $C_{i1}$ and $C_{i'1}$ anticommute for $i\neq i'$.
See \cref{noncon_app} for further discussion of noncontextual sets.

\section{Quasi-quantized model for a noncontextual Hamiltonian}
\label{model}

In this section we describe our main tool, a \emph{quasi-quantized} (or \emph{epistricted}) model for noncontextual Hamiltonians.
A quasi-quantized model is composed of a set of phase space points, called the \emph{ontic states} of the system, and a set of probability distributions over the ontic states, called the \emph{epistemic states} \cite{spekkens07a,spekkens16a}.

\subsection{Ontic states}
\label{phasespace}

If the set $\mathcal{S}$ is noncontextual, then consistent joint valuations for $\mathcal{S}$ exist.
However, in general not \emph{every} assignment of values to $\mathcal{S}$ is consistent.
This motivates the following:
\begin{definition}
\label{independentdef}
    An \emph{independent set} of Pauli operators contains no operator that can be written as a product of other commuting operators in the set.
\end{definition}
\noindent
In other words, a set is independent if it contains no operators whose values can be inferred from the values of other operators in the set, according to the notion of inference defined in \cref{noncontextual_intro}.
If a set of Pauli operators is independent, then \emph{every} joint valuation for the set is consistent.
Although the set $\mathcal{S}$ of Hamiltonian terms may be dependent, we can construct an independent set $\mathcal{R}$ such that $\overline{\mathcal{R}}=\overline{\mathcal{S}}$.
Since $\mathcal{R}$ is independent, the joint valuations for $\mathcal{R}$ are $\{+1,-1\}^{|\mathcal{R}|}$: these label the ontic states.

To obtain $\mathcal{R}$, we first construct
\begin{equation}
\label{gprime}
    G'\equiv\mathcal{Z}\cup\left(\bigcup_{i=1}^N\{A_{ij}~|~j=2,3,...,|C_i|\}\right).
\end{equation}
All operators in $G'$ commute universally in $\mathcal{S}$.
$G'$ will in general be dependent, but we may obtain an independent set $G\equiv\{G_i~|~i=1,2,...,|G|\}$ from $G'$ using the method described in \cite[\S10.5.7]{nielsen01} (summarized in \cref{independentcommutingset}). Operators in $G$ commute universally in $\mathcal{S}$ because $G\subseteq\overline{G'}$.

The independent set $\mathcal{R}$ is then given by
\begin{equation}
    \label{rdef}
    \mathcal{R}\equiv\{C_{i1}~|~i=1,2,...,N\}\cup G.
\end{equation}
The $C_{i1}$ are independent because no pair of them commutes, and since any product of operators in $G$ commutes with all operators in $\mathcal{S}$, the $C_{i1}$ are also independent of $G$.
Therefore, $\mathcal{R}$ is an independent set.
    
Each operator in $\mathcal{Z}$ and each $A_{ij}$ may be written as a product of operators in $G$ (see \cref{independentcommutingset}).
Therefore, since
\begin{equation}
    \mathcal{T}=\bigcup_{i=1}^N\{A_{ij}C_{i1}~|~j=1,2,...,N\},
\end{equation}
each operator in $\mathcal{T}$ is a product of one of the $C_{i1}$ and some set of operators in $G$.
Thus, each operator in $\mathcal{S}$ is a product of commuting operators in $\mathcal{R}$, so $\overline{\mathcal{R}}=\overline{\mathcal{S}}$.

The set $G$ has size at most $n-1$, because a set of $n$ independent, commuting Pauli operators forms a complete, commuting set of observables for $n$ qubits.
Therefore, if $G$ had size $n$ (or more), each set $G\cup\{C_{i1}\}$ would be a commuting set of size $n+1$ (or more), which could not be independent.
Thus $\mathcal{R}$ has size at most $N+n-1$.
As pointed out in \cite[\S IV.A]{raussendorf19b}, $N\le2m+1$ for $m\equiv n-|G|$, so $|\mathcal{R}|$ is in fact at most $2n+1$, which can be true only when $G$ (and thus $\mathcal{Z}$) is empty.
This will be important in \cref{simulation,approx_sect}.

Note that although $\mathcal{R}$ is not in general a subset of $\mathcal{S}$, it is a subset of $\overline{\mathcal{S}}$. Furthermore, since there is a bijection from the set of ontic states for $\mathcal{R}$ to the set of ontic states for $\overline{\mathcal{S}}$, $|\mathcal{R}|$ is unique for each $\overline{\mathcal{S}}$. Thus closure under inference permits construction of the independent generating set $\mathcal{R}$.

\subsection{Epistemic states}
\label{probdists}

Epistemic states are joint probability distributions over the ontic states, which complete our quasi-quantized model.
We write these joint probabilities as $P(c_1,...,c_N,g_1,g_2,...)$, where each $c_i,g_i$ is $\pm1$ and denotes the value assigned to $C_{i1}$ or $G_i$, respectively.

First consider a commuting Hamiltonian.
In this case, $\mathcal{R}=G$, and $\overline{\mathcal{S}}=\overline{\mathcal{R}}$ is the Abelian group generated by $\mathcal{R}$.
The observables may be simultaneously measured, so there is a one-to-one mapping between the ontic states and the simultaneous eigenstates of $\mathcal{S}$.
Thus the only constraint on the joint probabilities in this case is normalization.

Next consider the case where all observables pairwise anticommute:
\begin{lemma}
    \label{blochball}
    Let $\vec{A}=(A_1,A_2,...,A_N)$ be an anticommuting set of Pauli operators.
    For any unit vector $\vec{a}\in\mathbb{R}^N$, the operator $\sum_{i=1}^Na_iA_i$ has eigenvalues $\pm1$.
    From this it follows that for any state, $\sum_{i=1}^N\langle A_i\rangle^2\le1$.
\end{lemma}
\noindent
We prove \cref{blochball} in \cref{proofs}.

For a general noncontextual set $\mathcal{S}$, construct $\mathcal{R}$ as described in \cref{phasespace}.
The set $E=E(\mathcal{R})$ of epistemic states is then:
\begin{equation}
    \label{epistemicstates}
    E\equiv\left\{(\vec{q},\vec{r})\in\{\pm1\}^{|G|}\times\mathbb{R}^N~|~|\vec{r}|=1\right\}.
\end{equation}
The pairs $(\vec{q},\vec{r})$ define the joint probabilities as follows:
\begin{equation}
    \label{epistemictojoint}
    P_{(\vec{q},\vec{r})}(c_1,...,c_N,g_1,g_2,...)=\left(\prod_{j=1}^{|G|}\delta_{g_j,q_j}\right)\prod_{i=1}^N\frac{1}{2}|c_i+r_i|.
\end{equation}
We refer to both the joint probabilities and the vector pairs $(\vec{q},\vec{r})$ as epistemic states: they contain equivalent information.
    
In terms of $(\vec{q},\vec{r})$, the expectation values for $\mathcal{R}$ are given by
\begin{equation}
    \label{expvals}
    \begin{split}
        &\langle G_j\rangle_{(\vec{q},\vec{r})}=q_j,\\
        &\langle C_{i1}\rangle_{(\vec{q},\vec{r})}=r_i.
    \end{split}
\end{equation}
\begin{theorem}
\label{expvalsthm}
    For epistemic states $(\vec{q},\vec{r})$ as defined in \eqref{epistemicstates}, the joint probability distribution \eqref{epistemictojoint} is equivalent to the set of expectation values \eqref{expvals}.
\end{theorem}
\noindent
We prove \cref{expvalsthm} in \cref{proofs}.

The model \eqref{epistemicstates} is epistricted in the following sense: as in \cite{spekkens16a}, a state is represented by joint knowledge of a set of commuting observables.
For a given $(\vec{q},\vec{r})$, this set is $G$ together with the observable 
\begin{equation}
    \mathcal{A}(\vec{r})\equiv\sum_{i=1}^Nr_iC_{i1}
\end{equation}
(which has eigenvalues $\pm1$, by \cref{blochball}).
Note that since the $C_{i1}$ have expectation values $r_i$ as in \eqref{expvals}, $\mathcal{A}(\vec{r})$ has expectation value $1$, since $\vec{r}$ is a unit vector.
No probability distributions are allowed that represent more knowledge of the state than simultaneous values of $G$ and $\mathcal{A}(\vec{r})$.
Note that our model describes only pure states (as do the models in \cite{spekkens16a}).

From the expectation values \eqref{expvals} for $\mathcal{R}$, we can obtain expectation values for $\mathcal{S}$ as follows.
For $B\in\mathcal{Z}$, let $\mathcal{J}_B$ be the set of indices such that $B=\prod_{j\in\mathcal{J}_B}G_j$; then
\begin{equation}
    \label{expvalscomm}
    \langle B\rangle_{(\vec{q},\vec{r})}=\left\langle\prod_{j\in\mathcal{J}_B}G_j\right\rangle=\prod_{j\in\mathcal{J}_B}q_j,
\end{equation}
where the second equality follows because $G_j\mapsto q_j=\pm1$ for all $j$ (in other words, the state is a common eigenstate of the $G_j$ and of $B$).
Similarly, for $C_{i1}B\in\mathcal{T}$,
\begin{equation}
    \label{expvalsanticomm}
    \langle C_{i1}B\rangle_{(\vec{q},\vec{r})}=r_i\prod_{j\in\mathcal{J}_B}q_j.
\end{equation}

\begin{theorem}
\label{correspondence_thm}
The epistemic states \eqref{epistemicstates} give sets of expectation values that correspond to valid quantum states, and the set of quantum states described by the epistemic states includes an eigenbasis of any Hamiltonian whose Pauli terms are $\mathcal{S}$.
\end{theorem}

The proof of \cref{correspondence_thm} is given in \cref{proofs}.
Note that for any $(\vec{q},\vec{r})$, the expectation values \eqref{expvals} are produced by a simultaneous eigenstate of $G\cup\{\mathcal{A}(\vec{r})\}$.
For the second claim in \cref{correspondence_thm}, we show that there exists an eigenbasis for the Hamiltonian composed of common eigenstates of $G$ and $\mathcal{A}(\vec{r})$ for some $\vec{r}$.
That $\mathcal{A}(\vec{r})$ may be included is implied by:
\begin{lemma}
    \label{anticommexpthm}
    For $|\psi\rangle$ an eigenstate of the full Hamiltonian \eqref{ham}, the expectation values of the $C_{i1}$ satisfy $\sum_{i=1}^N\langle C_{i1}\rangle^2=1$.
\end{lemma}
\noindent
The proof of \cref{anticommexpthm} is given in \cref{proofs}.
In other words, the $\langle C_{i1}\rangle$ saturate the bound given in \cref{blochball}, for any energy eigenstate.
This means that every energy eigenvalue can be reproduced via the expectation values \eqref{expvalscomm} and \eqref{expvalsanticomm} for some setting of $(\vec{q},\vec{r})$.

We show in \cref{models_app} how for \emph{any} quantum state we may construct a joint probability distribution that reproduces the expectation values for $\overline{\mathcal{S}}$; however, to simulate noncontextual VQE it is only necessary to reproduce probabilities corresponding to eigenstates.

\section{Classical simulation of a noncontextual Hamiltonian}
\label{simulation}

\subsection{Classical objective function}
\label{objfn_sec}

Given the model described in \cref{model}, we now define a classical variational algorithm to simulate a noncontextual Hamiltonian.
In \eqref{ham}, each $A_{ij}$ and each $B$ is a product of operators in $G$, i.e., is an element of $\overline{G}$.
Therefore, we may replace $A_{ij}$ by $B$ and sum over all of $\overline{G}$, obtaining:
\begin{equation}
    \label{gennonconham}
    H=\sum_{B\in\overline{G}}\left(h_BB+\sum_{i=1}^Nh_{B,i}BC_{i1}\right),
\end{equation}
where the $h_{B,i}$ and $h_B$ are real coefficients.
Thus we can use \eqref{expvalscomm} and \eqref{expvalsanticomm} to write:
\begin{equation}
\label{objfn}
    \langle H\rangle_{(\vec{q},\vec{r})}
    =\sum_{B\in\overline{G}}\left(h_B+\sum_{i=1}^Nh_{B,i}r_i\right)\prod_{j\in\mathcal{J}_B}q_j.
\end{equation}
We may now treat \eqref{objfn} as a classical objective function.

This classical optimization problem will in general be hard.
Although a convex special case of \eqref{objfn} is obtained when we fix some set of values for the $q_j$ (or when $G$ is empty), in general the function is non-convex.
It is also in general frustrated, even if all terms commute, since in that case \eqref{objfn} becomes a linear combination of products of the $q_j$.
Thus, as discussed further in \cref{hamprob_sec}, we should not expect worst-cases of this optimization to be tractable, but they remain at worst classically hard.

\subsection{The noncontextual Hamiltonian problem}
\label{hamprob_sec}

The statement of the \textsc{noncontextual Hamiltonian problem} is as follows:
The inputs are a Hamiltonian $H$ of the form \eqref{gennonconham} with $\text{poly}(n)$ terms, together with a ``promise" that the lowest eigenvalue of $H$ is either greater than $b$ or less than $a$ for some $a<b\in\mathbb{R}$ such that $b-a>1/\text{poly}(n)$.
The goal is to return YES if the lowest eigenvalue is less than $a$.

A problem is in NP if for every YES instance, there exists a proof (or \emph{witness}) that is classically verifiable in polynomial time.
In our case, the witness for a YES instance is an epistemic state $(\vec{q},\vec{r})\in E$ (a vector with dimension at most $2n+1$, as discussed in \cref{phasespace}) satisfying $\langle H\rangle_{(\vec{q},\vec{r})}<a$.
By showing that such a witness can be efficiently verified, we prove the following in \cref{proofs}:
\begin{theorem}
\label{npproof}
    The \textsc{noncontextual Hamiltonian problem} is in NP.
\end{theorem}

The \textsc{diagonal local Hamiltonian problem} is NP-complete, as follows from \cite{barahona82a}; see also \cite{yan12a}.
This remains true even for 2-local diagonal Hamiltonians \cite{wocjan03a}, so since any 2-local diagonal Hamiltonian is noncontextual with $\text{poly}(n)$ Pauli terms, the \textsc{noncontextual Hamiltonian problem} is NP-complete as well.

\section{Approximation of general Hamiltonians by noncontextual Hamiltonians}
\label{approx_sect}

\begin{table*}[t]
  \begin{tabular}{ l c c c c c c c c c }
    Citation\hspace{1in} & \hspace{0.05in}System\hspace{0.05in} & \hspace{0.05in}$n$\hspace{0.05in} & \hspace{0.05in}$|\mathcal{S}_\text{full}|$\hspace{0.05in} & \hspace{0.05in}$|\mathcal{S}_\text{noncon}|$\hspace{0.05in} & \hspace{0.05in}$|\mathcal{R}|$\hspace{0.05in} & \hspace{0.05in}$\eps_\text{noncon}$\hspace{0.05in} & \hspace{0.05in}$\eps_\text{diag}$\hspace{0.05in} & \hspace{0.05in}$\eps_\text{expt}$\hspace{0.05in} & \begin{tabular}[c]{@{}c@{}}Expt. outperforms\\noncontextual?\end{tabular} \\
    \hline
    Peruzzo \emph{et al.}, 2014 \cite{peruzzo14a} & HeH$^+$& 2 & 9 & 5 & 3 & 0.21 & 4.1 & 4.1 & No \\
    \hline
    Hempel \emph{et al.}, 2018 \cite{hempel18a} & LiH & 3 & 13 & 9 & 4 & 0.56 & 0.56 & $\sim$80 & No \\
    \hline
    Kandala \emph{et al.}, 2017 \cite{kandala17a} & LiH & 4 & 99 & 23 & 5 & 4.2 & 9.3 & $\sim$30 & No \\
    \hline
    Kandala \emph{et al.}, 2017 \cite{kandala17a} & BeH$_2$ & 6 & 164 & 42 & 7 & 156 & 266 & $\sim$90 & Yes
  \end{tabular}
\caption{Contextual VQE experiments, as approximated by noncontextual and diagonal Hamiltonians. $n$ is the number of qubits. $|\mathcal{S}_\text{full}|$ is the number of terms in the full Hamiltonian, $|\mathcal{S}_\text{noncon}|$ is the number of terms in the noncontextual sub-Hamiltonian, and $|\mathcal{R}|$ is the number of parameters in an epistemic state (which is upper bounded by $2n+1$ for $n$ qubits). $\eps_\text{noncon}$ is the error in the noncontextual approximation, $\eps_\text{diag}$ is the error obtained by only keeping the diagonal terms in the Hamiltonian, and $\eps_\text{expt}$ is the error in the VQE experiment. Errors are in units of chemical accuracy, 0.0016Ha. Experimental errors preceded by $\sim$ were estimated from figures.\label{approximationtable}}
\end{table*}

Can we use our classical simulation technique as an approximation method for contextual Hamiltonians?
Since the model described in \cref{model} depends structurally on the Hamiltonian being noncontextual, we cannot apply it to a general Hamiltonian directly.
However, given a general Hamiltonian we can find a noncontextual subset of the terms, and take their (collective) ground state energy as an approximation of the true value.

Finding the largest noncontextual subset of terms is a generalization of the disjoint cliques problem \cite{jansen93a}, which is NP-complete.
However, to date VQE experiments have largely focused on Hamiltonians in which the total weight of the terms (in $l_1$ norm) is dominated by the diagonal terms. Therefore, as a heuristic we select terms from the full Hamiltonian greedily by coefficient magnitude while the set remains noncontextual, thus obtaining the diagonal terms together with some additional set, of relatively low weight. Given this noncontextual set, we construct $\mathcal{R}$ and minimize the resulting objective function \eqref{objfn} by brute-force search, since for few qubits $\mathcal{R}$ is small (see \cref{phasespace} and \cref{approximationtable}). For larger examples a classical optimization technique should replace this brute force search.

We applied this heuristic to contextual Hamiltonians that have been simulated in VQE experiments to date.
The results are given in Table~\ref{approximationtable}.
The best noncontextual sub-Hamiltonians we found, for each full Hamiltonian in Table~\ref{approximationtable}, are listed in \cref{nonconsubhams}.

For small Hamiltonians, our noncontextual approximation reached chemical accuracy. In all but one case the noncontextual approximation outperformed the approximation obtained by keeping only the diagonal terms. This is a natural point of comparison, since diagonal Hamiltonians constitute another common notion of classicality, and any set of diagonal Pauli operators is noncontextual.

Also, in all cases except for the BeH$_2$ simulation in \cite{kandala17a}, the noncontextual approximation reached better accuracy than the corresponding experiment.
We wish to be clear that this is not a criticism of these experiments, which were intended as demonstrations of methodology rather than as precise estimations.
However, what our noncontextual approximations show is that these experiments have not achieved sufficient accuracy to resolve intrinsically quantum behavior, i.e., the full-configuration correction to the noncontextual ground state energy.

Finally, the fact that the BeH$_2$ experiment does outperform our approximation indicates that, as we would expect, more terms in the Hamiltonian means more room for contextuality, and hence worse noncontextual approximations. Thus, we may hope that future experiments simulating larger Hamiltonians will reliably exceed this minimum standard for quantum behavior. On the other hand, better heuristics for identifying the noncontextual set may improve the noncontextual approximation.

\section{Discussion}
\label{discussion}

In the quantum approximate optimization algorithm (QAOA) \cite{farhi14a}, the Hamiltonian is diagonal (and thus noncontextual), because it encodes a classical problem, so our method simply recovers the diagonal entries.
We have shown that the \textsc{noncontextual Hamiltonian problem} is in NP.
Thus, the potential for quantum advantage in noncontextual VQE reduces to the same question that motivates QAOA: can we use cleverly-chosen and/or physically-motivated ansatze to generate otherwise hard-to-reach joint probability distributions, and thus efficiently converge to solutions of classically hard problems?

We demonstrated in \cref{approx_sect} that our model is applicable as an approximation method for contextual Hamiltonians, such as general electronic-structure Hamiltonians.
This technique provides a more stringent test for nonclassicality than that given Table I in \cite{kirby19a}, by demonstrating that some experiments with contextual Hamiltonians do not achieve sufficient accuracy to tell their results apart from those due to a noncontextual approximation.
Again, we wish to stress that this is not a criticism of these experiments, which have played seminal roles in the development of quantum simulation techniques, but only a means by which we may try to identify intrinsically quantum behavior.

Finally, in addition to serving as a benchmark for quantum experiments, our simulation technique may be useful as a new approximation method in its own right.
Also, it may be possible to extend our criterion for noncontextuality to other Hamiltonian decompositions besides Pauli decomposition, thus improving the capacity of our simulation algorithm.
We leave the full exploration of these possibilities for future work.

\begin{acknowledgements}
W. M. K. acknowledges support from the National Science Foundation, Grant No. DGE-1842474.
P. J. L. acknowledges support from the National Science Foundation, Grant No. PHY-1720395, and from Google Inc.
This work was supported by the National Science Foundation STAQ project (PHY-1818914).
\end{acknowledgements}

\bibliography{references.bib}

\appendix

\section{Proofs}
\label[appendix]{proofs}

\noindent
\textbf{Lemma~\ref{blochball}}
\emph{
    Let $\vec{A}=(A_1,A_2,...,A_N)$ be an anticommuting set of Pauli operators.
    For any unit vector $\vec{a}\in\mathbb{R}^N$, the operator $\vec{a}\cdot\vec{A}\equiv\sum_{i=1}^Na_iA_i$ has eigenvalues $\pm1$.
    From this it follows that for any state, $\sum_{i=1}^N\langle A_i\rangle^2\le1$.
}
\begin{proof}~
    This theorem consists of two claims:
    
    For the first claim (that the operator $\vec{a}\cdot\vec{A}$ has eigenvalues $\pm1$), note that
    \begin{align}
        (\vec{a}\cdot\vec{A})^2&=\sum_ia_i^2A_i^2+\sum_{i>j}a_ia_j\{A_i,A_j\}\nonumber\\
        &=\sum_ia_i^2\mathds{1}=|\vec{a}|^2\mathds{1}=\mathds{1},
    \end{align}
    where the second equality follows because the Pauli operators $A_i$ are self-inverse, and by assumption $\{A_i,A_j\}=0$ for $i>j$.
    Thus, $\vec{a}\cdot\vec{A}$ is self-inverse, so its eigenvalues are $\pm1$.
    This completes the proof of the first claim.
    
    For the second claim (that $\sum_{i=1}^N\langle A_i\rangle^2\le1$), note that for any state represented as a density operator $\rho$, we may write
    \begin{equation}
        \rho=\frac{1}{2^n}\left(\mathds{1}+\vec{b}\cdot\vec{A}+\cdots\right),
    \end{equation}
    for some $\vec{b}\in\mathbb{R}^N$, where the ellipsis indicates the presence in general of additional Pauli terms.
    The expectation value of any $A_i$ is then
    \begin{equation}
        \label{pauliexp}
        \langle A_i\rangle=\text{Tr}(A_i\rho)=b_i,
    \end{equation}
    since the Pauli operators are Hilbert-Schmidt orthogonal and self-inverse.
        
    Assume that $|\vec{b}|>0$ (if $|\vec{b}|=0$ then the theorem holds, by \eqref{pauliexp}.)
    Let $\vec{a}\equiv\frac{\vec{b}}{|\vec{b}|}$, so that $|\vec{a}|=1$.
    By the first claim in this lemma, the observable $\vec{a}\cdot\vec{A}$ has eigenvalues $\pm1$.
    Therefore, $|\langle\vec{a}\cdot\vec{A}\rangle|\le1$, so we have
    \begin{align}
        1\ge&|\langle\vec{a}\cdot\vec{A}\rangle|=|\text{Tr}(\vec{a}\cdot\vec{A}\rho)|\nonumber\\
        &=\left|\frac{1}{2^n}\text{Tr}\left(\vec{a}\cdot\vec{A}(\vec{b}\cdot\vec{A})\right)\right|=|\vec{a}\cdot\vec{b}|,
    \end{align}
    where (as above) the steps in the second line follow because the Pauli operators are Hilbert-Schmidt orthogonal and self-inverse.
    Therefore,
    \begin{equation}
        1\ge|\vec{a}\cdot\vec{b}|^2=\frac{|\vec{b}|^4}{|\vec{b}|^2}=|\vec{b}|^2=\sum_{i=1}^Nb_i^2=\sum_{i=1}^N\langle A_i\rangle^2
    \end{equation}
    (since by assumption $|\vec{b}|>0$.)
    This completes the proof of the second claim.
\end{proof}
~

\noindent
\textbf{Theorem~\ref{expvalsthm}}
\emph{
    For epistemic states $(\vec{q},\vec{r})$ as defined in \eqref{epistemicstates}, the joint probability distribution \eqref{epistemictojoint} is equivalent to the set of expectation values \eqref{expvals}.
}
\begin{proof}~
\begin{enumerate}
    \item We first prove the reverse implication: assume that \eqref{expvals} holds.
    We reproduce \eqref{expvals} here for convenience:
    \begin{equation}
        \label{expvals_app}
        \begin{split}
            &\langle G_j\rangle_{(\vec{q},\vec{r})}=q_j,\\
            &\langle C_{i1}\rangle_{(\vec{q},\vec{r})}=r_i.
        \end{split}
    \end{equation}
    That these lead to the joint probabilities given by \eqref{epistemictojoint} essentially follows from the discussion following \eqref{expvals}, but we will fill in the details.

    Since $\langle G_j\rangle=q_j=\pm1$ for each $j$ (as in \eqref{epistemicstates}), the $G_j$ have definite values.
    The values of the $G_j$ in the ontic state $(c_1,...,c_N,g_1,g_2,...)$ are the $g_j$, so $P(c_1,...,c_N,g_1,g_2,...)$ can be nonzero only when $q_j=g_j$ for each $j$.
    If this holds for each $j$, then $P(c_1,...,c_N,g_1,g_2,...)$ is just the product over $i=1,2,...,N$ of the probabilities of obtaining outcome $C_{i1}~\mapsto~c_i$ given the expectation value $\langle C_{i1}\rangle=r_i$: these probabilities are given by
    \begin{equation}
        \frac{1}{2}|c_i+r_i|
    \end{equation}
    for each $i$.
    Taking the joint probability to be the product of these works only because the $C_{i1}$ do not commute, and thus cannot be correlated.

    The condition due to the $G_j$ thus gives a factor of
    \begin{equation}
        \prod_{j=1}^{|G|}\delta_{g_j,q_j},
    \end{equation}
    and the condition due to the $C_{i1}$ gives a factor of
    \begin{equation}
        \prod_{i=1}^N\frac{1}{2}|c_i+r_i|;
    \end{equation}
   \eqref{epistemictojoint} is simply the product of these.

    \item Now assume that \eqref{epistemictojoint} holds.
    We reproduce \eqref{epistemictojoint} here for convenience:
    \begin{equation}
        \label{epistemictojoint_app}
        P(c_1,...,c_N,g_1,g_2,...)=\left(\prod_{j=1}^{|G|}\delta_{g_j,q_j}\right)\left(\prod_{i=1}^N\frac{1}{2}|c_i+r_i|\right).
    \end{equation}
    The probabilities for the outcomes of each $G_j$ and each $C_{i1}$ should be obtained as marginals of \eqref{epistemictojoint_app}.
    If $p_{G_j}$ denotes the probability of obtaining the outcome $G_j~\mapsto+1$, then
    \begin{equation}
    \label{gprob}
    \begin{split}
        p_{G_j}&=\sum_{\substack{c_i,g_k=\pm1,\\\forall~k\neq j}}P(c_1,...,c_N,g_1,g_2,...)\bigg|_{g_j=1}\\
        &=\sum_{\substack{c_i,g_k=\pm1,\\\forall~k\neq j}}\left(\prod_{l=1}^{|G|}\delta_{g_l,q_l}\right)\bigg|_{g_j=1}\left(\prod_{m=1}^N\frac{1}{2}|c_m+r_m|\right)\\
        &=\delta_{q_j,1}\sum_{c_i=\pm1}\left(\prod_{m=1}^N\frac{1}{2}|c_m+r_m|\right)\\
        &=\frac{1}{2}(q_j+1)\left(\prod_{m=1}^N\frac{1}{2}(|r_m+1|+|r_m-1|)\right)\\
        &=\frac{1}{2}(q_j+1),
    \end{split}
    \end{equation}
    where the fourth equality follows because for $r_m\in[-1,1]$,
    \begin{equation}
        \label{rnormalizer}
        |r_m+1|+|r_m-1|=2.
    \end{equation}
    Similarly,
    \begin{equation}
    \label{cprob}
    \begin{split}
        p_{C_{i1}}&=\sum_{\substack{c_k,g_j=\pm1,\\\forall~k\neq i}}P(c_1,...,c_N,g_1,g_2,...)\bigg|_{c_i=1}\\
        &=\sum_{\substack{c_k,g_j=\pm1,\\\forall~k\neq i}}\left(\prod_{l=1}^{|G|}\delta_{g_l,q_l}\right)\left(\prod_{m=1}^N\frac{1}{2}|c_m+r_m|\right)\bigg|_{c_i=1}\\
        &=\sum_{\substack{c_k=\pm1,\\\forall~k\neq i}}\left(\prod_{m=1}^N\frac{1}{2}|c_m+r_m|\right)\bigg|_{c_i=1}\\
        &=\frac{1}{2}|r_i+1|=\frac{1}{2}(r_i+1),
    \end{split}
    \end{equation}
    again using \eqref{rnormalizer}.
    From \eqref{gprob} and \eqref{cprob} we can obtain the expectation values for $G_j$ and $C_{i1}$:
    \begin{equation}
        \begin{split}
            &\langle G_j\rangle_{(\vec{q},\vec{r})}=2p_{G_j}-1=q_j,\\
            &\langle C_{i1}\rangle_{(\vec{q},\vec{r})}=2p_{C_{i1}}-1=r_i,
        \end{split}
    \end{equation}
    which is \eqref{expvals} (and \eqref{expvals_app}, above), as desired.
\end{enumerate}
\end{proof}
~

\noindent
\textbf{Lemma~\ref{anticommexpthm}}
\emph{
    For $|\psi\rangle$ an eigenstate of the full Hamiltonian \eqref{ham}, the expectation values of the $C_{i1}$ satisfy
    \begin{equation}
        \sum_{i=1}^N\langle C_{i1}\rangle^2=1.
    \end{equation}
}
\begin{proof}
    As discussed in the main text, we may simultaneously diagonalize the $G_j$.
    The resulting form for the full Hamiltonian $H$ will be block-diagonal, with each block corresponding to a common eigenspace of the $G_j$, since the $C_{i1}$ commute with the $G_j$ and therefore do not mix their common eigenspaces.
    Thus, the eigenstates of the full Hamiltonian are common eigenstates of $G$.
    
    Given this, if $|\psi\rangle$ is an eigenstate of the full Hamiltonian, then $|\psi\rangle$ is also an eigenstate of any $B\in\overline{G}$, a product of some subset of $G$.
    Therefore,
    \begin{equation}
        BC_{i1}|\psi\rangle=C_{i1}B|\psi\rangle=\lambda_BC_{i1}|\psi\rangle,
    \end{equation}
    where $\lambda_B=\pm1$ is the eigenvalue of $|\psi\rangle$ for the operator $B$.
    Thus, we may rewrite the full Hamiltonian \eqref{ham}, acting on $|\psi\rangle$, as
    \begin{equation}
        \label{reducedHam}
        H|\psi\rangle=\left(h'_0+\sum_{i=1}^Nh'_iC_{i1}\right)|\psi\rangle,
    \end{equation}
    for coefficients $h'_i$ defined by
    \begin{equation}
    \label{reducedHamCoeffs}
    \begin{split}
        &h'_0\equiv\sum_{B\in\mathcal{Z}}h_B\lambda_B,\\
        &h'_i\equiv\sum_{B\in\overline{G}}h_{B,i}\lambda_B\quad\text{for $i>0$},
    \end{split}
    \end{equation}
    where $h_{B,i}$ is the coefficient of $BC_{i1}$ in the Hamiltonian.
    Since $|\psi\rangle$ is an eigenstate of $H$, by \eqref{reducedHam} it must also be an eigenstate of
    \begin{equation}
        \label{subHam}
        \sum_{i=1}^Nh'_iC_{i1}=h\sum_{i=1}^N\tilde{h}_iC_{i1},
    \end{equation}
    the non-identity terms in \eqref{reducedHam}, for $h\equiv\sqrt{\sum_{i=1}^N(h'_i)^2}$, so that the $\tilde{h}_i\equiv h'_i/h$ satisfy
    \begin{equation}
        \label{unitcoeffs}
        \sum_{i=1}^N\tilde{h}_i^2=1.
    \end{equation}
    Thus by \cref{blochball}, the operator $\sum_{i=1}^N\tilde{h}_iC_{i1}$ has eigenvalues $\pm1$.
    Since $|\psi\rangle$ is an eigenstate of the operator given in \eqref{subHam}, it is an eigenstate of $\sum_{i=1}^N\tilde{h}_iC_{i1}$:
    \begin{equation}
        \left(\sum_{i=1}^N\tilde{h}_iC_{i1}\right)|\psi\rangle=\pm|\psi\rangle.
    \end{equation}
    Therefore, if $\langle\cdot\rangle$ denotes expectation value with respect to $|\psi\rangle$,
    \begin{equation}
        \label{expvalcond}
        \langle\psi|\left(\sum_{i=1}^N\tilde{h}_i C_{i1}\right)|\psi\rangle=\sum_{i=1}^N\tilde{h}_i\langle C_{i1}\rangle=\pm1.
    \end{equation}
    We know by \cref{blochball} that
    \begin{equation}
        \sum_{i=1}^N\langle C_{i1}\rangle^2\le1,
    \end{equation}
    and by construction $\sum_{i=1}^N\tilde{h}_i^2=1$.
    Thus the only way \eqref{expvalcond} can be satisfied is if
    \begin{equation}
        \langle C_{i1}\rangle=s\tilde{h}_i
    \end{equation}
    for all $i=1,2,...,N$, for fixed $s=\pm1$.
    Therefore,
    \begin{equation}
        \sum_{i=1}^N\langle C_{i1}\rangle^2=\sum_{i=1}^N\tilde{h}_i^2=1.
    \end{equation}
    Since this holds for any $|\psi\rangle$ in the eigenbasis of $H$, it holds for any eigenstate of $H$.
\end{proof}
~

\noindent
\textbf{Theorem~\ref{correspondence_thm}}
\emph{
The epistemic states \eqref{epistemicstates} give sets of expectation values that correspond to valid quantum states, and the set of quantum states described by the epistemic states includes the eigenstates of any Hamiltonian whose Pauli terms are $\mathcal{S}$.
}
\begin{proof}
The theorem consists of two claims.
For the first, note that for any epistemic state $(\vec{q},\vec{r})$, there exists a simultaneous eigenstate of the operators $G_j$, with eigenvalue $q_j=\pm1$ for each $G_j$, and of $\mathcal{A}(\vec{r})\equiv\sum_{i=1}^Nr_iC_{i1}$, with eigenvalue $1$.
For this state, the expectation value of each $G_j$ is $q_j$, and the expectation value of each $C_{i1}$ is $r_i$, as noted in \cref{probdists}.
Thus every epistemic state $(\vec{q},\vec{r})$ corresponds to a valid quantum state, proving the first claim.

For the second claim, consider first the universally-commuting operators: $\mathcal{Z}$, which are generated by $G$.
Since these may in principle be simultaneously diagonalized, the common eigenstates of $G$ (which are the common eigenstates for all of $\mathcal{Z}$) are a complete set of eigenstates for any Hamiltonian whose terms are a linear combination of $\mathcal{Z}$.

Now suppose we add to such a Hamiltonian a linear combination of the terms in $\mathcal{T}$, each of which is a product of operators in $G$ with one of the $C_{i1}$.
As in the previous paragraph, we may simultaneously diagonalize $G$: the resulting form for the full Hamiltonian will be block-diagonal, with each block corresponding to a common eigenspace of the $G_j$.
Within each block, the Hamiltonian will take the form of some linear combination of the $C_{i1}$, obtained by replacing the $G_j$ by their eigenvalues for the current block, as in \eqref{reducedHam} and \eqref{reducedHamCoeffs}.
Since the $C_{i1}$ commute with $G$, we may diagonalize the linear combination of the $C_{i1}$ within each block (i.e., the $C_{i1}$ do not mix the common eigenspaces of $G$).
Thus, it is still the case that we may take the eigenstates of the full Hamiltonian to be common eigenstates of $G$.
This justifies the condition on $\vec{q}$ (the first set of values in our epistemic states) given in \eqref{epistemicstates}, namely, $q_j=\pm1$, since these correspond to the expectation values of the $G_j\in G$.

Regarding the condition on $\vec{r}$ (the second set of values in our epistemic state) in \eqref{epistemicstates}, we draw on \cref{anticommexpthm}.
In \eqref{epistemicstates}, the condition on $\vec{r}$ is that it be a unit vector, so since its components $r_i$ give the expectation values for the $C_{i1}$ as in \eqref{expvals}, \cref{anticommexpthm} proves that the set of expectation values thus described includes all eigenvalues of the Hamiltonian.
\end{proof}
~

\noindent
\textbf{Theorem~\ref{npproof}}
\emph{
The \textsc{noncontextual Hamiltonian problem} is in NP.
}
\begin{proof}
We demonstrated in \cref{objfn_sec} that by varying over all epistemic states, the expected energy given in \eqref{objfn} varies over all eigenvalues of the Hamiltonian.
Therefore, if the Hamiltonian possesses an eigenvalue $\lambda<a$, there is some epistemic state $(\vec{q},\vec{r})$ such that $\langle H\rangle_{(\vec{q},\vec{r})}=\lambda$.
This epistemic state serves as a proof of the YES instance, as long as given the Hamiltonian as in \eqref{gennonconham} we can classically evaluate the objective function in \eqref{objfn} in polynomial time.

To show this, we use the fact that by assumption $|\mathcal{S}|$ (the number of terms in the Hamiltonian) is polynomial in $n$.
Thus, the total number of nonzero coefficients $h_B$ and $h_{B,i}$ is $\text{poly}(n)$; these are given in the statement of the problem instance.
Which coefficients are nonzero also determines the terms in the sum in \eqref{objfn} (and upper bounds their number).
The only remaining components of \eqref{objfn} to be evaluated are the sets of indices $\mathcal{J}_B$: we have one such set for each term, defined to satisfy $B=\prod_{G\in\mathcal{J}_B}G$.
The $\mathcal{J}_B$ are obtained directly from the standard method used to construct $G$ (described in \cref{independentcommutingset}), the entirety of which requires $\text{poly}(n)$ classical operations given that $|\mathcal{S}|=\text{poly}(n)$.
Thus, given a Hamiltonian of the form \eqref{gennonconham} and a witness in the form of an epistemic state $(\vec{q},\vec{r})$, we can use \eqref{objfn} to verify that $\langle H\rangle_{(\vec{q},\vec{r})}<a$, in $\text{poly}(n)$ classical operations.
\end{proof}

\section{Noncontextual Sets of Pauli Operators}
\label[appendix]{noncon_app}

As noted in \cref{noncontextual_intro}, the key concept in reasoning about noncontextuality is inference among value assignments to Pauli operators.
This stems from the following property that we demand of a noncontextual ontological model for a set of Pauli observables: in any set of measurements we can perform simultaneously, the values we obtain from the noncontextual model must agree with those required by the full formalism of quantum mechanics.
In other words, if $A$ and $B$ are commuting Pauli operators, then by measuring them simultaneously we can predict with certainty the result of measuring their product $AB$; therefore, in an assignment of values to $\{A,B,AB\}$, the value assigned to $AB$ must be the product of the values assigned to $A$ and $B$.

This motivates the closure under inference $\overline{\mathcal{S}}$ of a set $\mathcal{S}$ of Pauli operators, as defined in \cref{noncontextual_intro}: it is the set of Pauli operators whose values are determined by a value assignment to $\mathcal{S}$ \cite{kirby19a}.
The Jordan product $\frac{1}{2}\{\cdot,\cdot\}$ provides another definition of $\overline{\mathcal{S}}$.
Since a pair of Pauli operators $A,B$ either commute or anticommute,
\begin{equation}
    \frac{1}{2}\{A,B\}=
    \begin{cases}
        AB\text{ if }[A,B]=0,\\
        0\text{ otherwise}.
    \end{cases}
\end{equation}
Thus, $\overline{\mathcal{S}}$ is the closure of $\mathcal{S}$ under the Jordan product.

A set of Pauli operators is noncontextual if ontic states for it exist that are self-consistent.
We check consistency of an ontic state by making sure that in the ontic state it induces for the closure under inference, none of the inferences are violated: in other words, the value assigned each each product $AB$ of a commuting pair $A,B$ is the product of the values assigned to the pair.

Since by definition an ontic state (assignment of values) for $\mathcal{S}$ extends to an ontic state for $\overline{\mathcal{S}}$, and the ontic state for $\overline{\mathcal{S}}$ includes the ontic state for $\mathcal{S}$, any pair of noncontextual sets of Pauli operators that share the same closure under inference have equivalent ontic states that may derived from each other.
Thus, given an arbitrary noncontextual set of Pauli operators, we may transform it into a noncontextual set with a standard form that shares the same closure under inference.

The first move we can make in doing this is, given a commuting pair $A,B$, so replace $B$ by $AB$.
Since Pauli operators are self-inverse, $B$ is given by the product of $A$ and $AB$, so from a value assignment to $A$ and $AB$ we can infer the value assignment to $B$, and thus recover the original set.
The second move we can make is to remove any $C$ that is the product of some commuting $A,B$ that are also in the set.
In obtaining the independent set $\mathcal{R}$ that we used to define our ontic states in \cref{phasespace}, we simply made a sequence of such moves.

The Jordan product is again useful in describing independent sets according to \cref{independentdef}.
If $\mathcal{S}$ commutes, then this definition reduces to that in \cite[\S10.5.1]{nielsen01}.
For the general noncontextual case, we would like to relate \cref{independentdef} to the notion of independence that applies to sets of generators of Abelian groups, where it means that removing any of the generators reduces the size of the generated group.
Since the Jordan product is commutative, \cref{independentdef} nearly reproduces the usual notion of independence for an Abelian group taking the group operation to be the Jordan product, but the Jordan product is not associative.
This distinction aside, $\overline{\mathcal{S}}$ is the closure of $\mathcal{S}$ under the Jordan product, and correspondingly, a subset of $\overline{\mathcal{S}}$ that is independent in our sense is also independent as a generating set under the Jordan product.

Thus, given an arbitrary noncontextual set $\mathcal{S}$, $\mathcal{R}$ as constructed in \cref{phasespace} is an independent generating set for $\overline{\mathcal{S}}$ under the Jordan product.
The construction we gave in the main text may therefore be summarized: a set of Pauli operators is noncontextual if and only if the independent generating sets for its closure under inference are composed of a set of universally-commuting operators ($G$), and a set of pairwise anticommuting operators $\{C_{i1}~|~i=1,2,...,N\}$.

\section{Obtaining an independent set of Pauli measurements from a commuting set}
\label[appendix]{independentcommutingset}

Given a set $G'$ of commuting Pauli operators on $n$ qubits, we wish to obtain an independent commuting set $G$ such that every operator in $G'$ is a product of operators in $G$.
Since $G$ commutes, $G$ is independent if and only if no operator in $G$ is a product of other operators in $G$ (see the discussion following \cref{independentdef}.)
Finding $G$ given $G'$ is a standard procedure, and a method for performing it is given by Nielsen and Chuang in \cite[\S10.5.7]{nielsen01}.
We summarize the method here in terms of the language used in this work.

The method to calculate $G$ from $G'$ is a multiplicative variant of Gaussian elimination.
Let $G'=\{g'_1,g'_2,...,g'_m\}$, and write each $g'_i$ as
\begin{equation}
    g'_i=h_{i1}\otimes h_{i2}\otimes\cdots\otimes h_{in},
\end{equation}
where each $h_{ij}$ is a single-qubit Pauli operator (including the identity).
We may then express $G'$ in an array as:
\begin{equation}
    \label{initialmatrix}
    G'=
    \begin{matrix}
        h_{11}&h_{12}&h_{13}&\cdots&h_{1n},\\
        h_{21}&h_{22}&h_{23}&\cdots&h_{2n},\\
        h_{31}&h_{32}&h_{33}&\cdots&h_{3n},\\
        &&&\vdots&\\
        h_{m1}&h_{m2}&h_{m3}&\cdots&h_{mn},\\
    \end{matrix}
\end{equation}
where we have suppressed the tensor product symbols.

Let $h$ denote the matrix whose entries are $h_{ij}$, the single-qubit Pauli operators appearing in \eqref{initialmatrix}, augmented by a vector $\vec{s}$ whose entries $s_i$ are the signs associated to each row in $h_i$ (initially these are all $+1$).
We first describe a procedure that, given such a matrix $h$, transforms it to a matrix in which at most two entries in the first column are non-identity:
\begin{enumerate}
    \item $h_{11},h_{21},...,h_{m1}$ are the entries in the first column of $h$. If
    \begin{equation}
        h_{11}=h_{21}=\cdots=h_{m1}=I,
    \end{equation}
    then we are already done.
    
    \item If for some $k$, $h_{k1}=X$, then for each $i\neq k$, if $h_{i1}=X$, multiply row $i$ by row $k$.
    This corresponds to multiplying $g'_i$ by $g'_k$, or the following mapping on the entries in $h$:
    \begin{equation}
    \begin{split}
        h_{i1}~&\mapsto~h_{i1}h_{k1}=X^2=I,\\
        h_{ij}~&\mapsto~h_{ij}h_{kj},~\forall j>1.
    \end{split}
    \end{equation}
    Thus when we have completed this step for each $i\neq k$, there will be no $X$s in the first column except in row $k$.
    
    \item If for some $l$, $h_{l1}=Z$, then for each $i\neq l$, if $h_{i1}=Z$, multiply row $i$ by row $l$.
    As in step 2, this corresponds to multiplying $g'_i$ by $g'_l$, so when we have completed this step for each $i\neq l$, there will be no $Z$s in the first column except in row $l$.
    
    \item If there is both an $X$ and a $Z$ in the first column (in rows $k$ and $l$, respectively), then for each $i\neq k,l$, if $h_{i1}=Y$, multiply row $i$ by row $k$ and row $l$.
    This corresponds to multiplying $g'_i$ by $g'_kg'_l$, or the following mapping on the entries in $h$:
    \begin{equation}
    \begin{split}
        h_{i1}~&\mapsto~h_{i1}h_{k1}h_{l1}=YXZ=-iI,\\
        h_{ij}~&\mapsto~h_{ij}h_{kj}h_{lj},~\forall j>1.
    \end{split}
    \end{equation}
    Thus when we have completed this step for each $i\neq l$, there will be no $Y$s in the first column at all (if both $X$ and $Z$ are present in the first column).
    
    \item If $X$ and $Z$ are not both present in the first column, and if for some $m$, $h_{m1}=Y$, then for each $i\neq m$, if $h_{i1}=Y$, multiply row $i$ by row $m$.
    As in step 2, this corresponds to multiplying $g'_i$ by $g'_m$, so when we have completed this step for each $i\neq m$, there will be no $Y$s in the first column except in row $m$.
\end{enumerate}

Note that since we know that the full Pauli operators corresponding to the rows commute, the total phase obtained in multiplying any row by any other row must be $\pm1$.
We record this by multiplying $s_i$ by this sign, when the multiplication takes place in the $i$th row.

Call this procedure $\textsc{ReduceFirstCol}(h)$.
When we have completed the mapping, the transformed matrix $h$ will contain at most two rows in which the first entry is non-identity (and the first entries in those two rows will be different).

Our full procedure to obtain $G$ is then as follows:
\begin{enumerate}
    \item Let $h$ be the full matrix \eqref{initialmatrix}.
    $h'$ will be a submatrix of $h$ that is updated at each iteration; let $h'$ initially be equal to $h$.
    
    \item Perform the mapping
    \begin{equation}
        h'~\mapsto~\textsc{ReduceFirstCol}(h').
    \end{equation}
    
    \item For each of the (up to) two rows in $h'$ in which the first entry is non-identity, put the corresponding Pauli operators in $G$.
    Then let the new $h'$ be the submatrix obtained from the current $h'$ by removing these rows and the first column.
    
    \item If any non-identity entries remain in $h'$, return to step 2.
    If not, then $G$ is complete.
\end{enumerate}
When this procedure is complete, at most two rows in $h$ will have a non-identity first entry, at most two others will have a non-identity second entry, at most two others will have a non-identity third entry, and so forth.
In other words, under some reordering of the rows, $h$ will have the form:
\begin{equation}
    \label{finalmatrix}
    \begin{matrix}
        &P_1&h'_{12}&h'_{13}&h'_{14}&h'_{15}&\cdots&h'_{1n},&\\
        [&P_2&h'_{22}&h'_{23}&h'_{24}&h'_{25}&\cdots&h'_{2n},&]\\
        &I&P_3&h'_{33}&h'_{34}&h'_{35}&\cdots&h'_{3n},&\\
        [&I&P_4&h'_{43}&h'_{44}&h'_{45}&\cdots&h'_{4n},&]\\
        &I&I&P_5&h'_{54}&h'_{55}&\cdots&h'_{5n},&\\
        [&I&I&P_6&h'_{64}&h'_{65}&\cdots&h'_{6n},&]\\
        &&&&\vdots&&&&\\
        [&I&I&I&I&I&\cdots&I,&]
    \end{matrix}
\end{equation}
where the bracketed rows may or may not appear, and in cases where they do appear $P_2\neq P_1$, $P_4\neq P_3$, $P_6\neq P_5$ and so forth.
Note that in general there may be a collection of rows at the bottom of the matrix that have been reduced entirely to the identity.
$G$ will be given by the non-identity rows in \eqref{finalmatrix}.

To see that $G$ is independent, consider some particular non-identity row $i$ (representing an operator $g_i\in G$).
Row $i$ cannot be written as a product of rows below it, since if $h_{ij}$ is row $i$'s first non-identity entry, $h_{i'j}=I$ for all $i'=i+2,i+3,...,m$.
The entry $h_{(i+1)j}$ (immediately below $h_{ij}$) might not be the identity, but it cannot be equal to $h_{ij}$, so no product of the entries below $h_{ij}$ can be equal to $h_{ij}$.
Thus $g_i$ cannot be written as the product of any subset of the $g_j$ for $j>i$.

But this means that $g_i$ cannot be written as a product of any subset of the other $g_j$ for $j\neq i$, as we can prove by contradiction: suppose $g_i$ could be written as a product
\begin{equation}
    \label{contradiction1}
    g_i=\prod_{j\in\mathcal{J}}g_j,
\end{equation}
where $\mathcal{J}$ is some set of indices not including $i$.
If $i<j$ for all $j\in\mathcal{J}$, then this violates the condition in the previous paragraph directly.
Otherwise, choose the least $j\in\mathcal{J}$: call it $j'$.
Then since the Pauli operators are self-inverse, and all of the $g_k$ commute, we can rearrange \eqref{contradiction1} as
\begin{equation}
    \label{contradiction2}
    g_{j'}=g_i\prod_{j\in\mathcal{J}\setminus\{j'\}}g_j,
\end{equation}
which violates the condition in the previous paragraph since $j'<i,j$ for all $j\in\mathcal{J}\setminus\{j'\}$.
Thus, all of the $g_i$ are independent.

We wish to be able to recover the original set of operators $G'$ from the new independent set $G$.
To accomplish this, we simply need to keep track of the row multiplications carried out in each step above.
Since the Pauli operators are self-inverse, we can reconstruct each row $i$ in $G'$ as the product of the corresponding row in $G$ and the rows it is multiplied by in the procedure (including in the multiplication the corresponding signs in $\vec{s}$).
Thus we can write each operator in $G'$ as a product of operators in $G$.

\section{More general quasi-quantized models}
\label[appendix]{models_app}

In \cref{phasespace}, we showed how to construct the ontic states (phase space points) for a noncontextual set of observables.
The key step is obtaining the set $\mathcal{R}=\{C_{i1}~|~i=1,2,...,N\}\cup G$, where
$\{C_{i1}~|~i=1,2,...,N\}$
is a completely anticommuting set,
$G=\{G_i~|~i=1,2,...,|G|\}$
is an independent commuting set whose elements also commute with each $C_{i1}$, and every operator in $\mathcal{S}$ may be written as a product of commuting operators in $\mathcal{R}$.
Specifically, each universally commuting operator in $\mathcal{S}$ (these compose the set $\mathcal{Z}$) may be written as a product of operators in $G$, and every operator in one of the cliques $C_i\subset\mathcal{T}$ may be written as a product of operators in $G$ with the single operator $C_{i1}$.

The joint outcome assignments for $\mathcal{R}$ label the phase space points (each of which defines a joint outcome assignment for all of $\mathcal{S}$), so they also label the joint probabilities associated to each phase space point.
As in \cref{probdists}, we write these joint probabilities as
\begin{equation}
    \label{jointprobs_app}
    P(c_1,...,c_N,g_1,g_2,...),
\end{equation}
where each $c_i,g_i$ is $\pm1$ and denotes the outcome assigned to $C_{i1}$ or $G_i$, respectively.

In \cref{probdists}, we presented a quasi-quantized model that gives a set of joint probabilities $P$ (by way of a set of expectation values) that is sufficiently general to reproduce the expectation values associated to any eigenstate of the Hamiltonian.
However, the set of ontic states in principle admits broader sets of probability distributions, which we discuss in this appendix.

The probabilities for outcome $+1$ for each operator in $\mathcal{S}$ should be obtained as marginals of the joint probability distribution \eqref{jointprobs_app}.
Consider an operator in $\mathcal{S}$ that is written as
\begin{equation}
    \label{operatorprod}
    B\text{ or }C_{i1}B,
\end{equation}
for $B$ a product of operators in $G$ (which, it is understood, may be the identity --- the product of no operators in $G$).
The corresponding marginal is the sum of the joint probabilities for outcomes such that the product outcome from \eqref{operatorprod} is $+1$.
In other words, if $p_{O}$ denotes the probability of obtaining outcome $+1$ upon measurement of the operator $O$, then
\begin{align}
    &p_B=\sum_{\substack{c_i,g_j=\pm1,\\\text{s.t. }B\thinspace\mapsto1}}P(c_1,...,c_N,g_1,g_2,...),\label{commops}\\
    &p_{C_{i1}B}=\sum_{\substack{c_i,g_j=\pm1,\\\text{s.t. }C_{i1}B\thinspace\mapsto1}}P(c_1,...,c_N,g_1,g_2,...).\label{anticommops}
\end{align}
Exactly half of the joint probabilities will appear in each sum \eqref{commops} and \eqref{anticommops}.
This is apparent when we note that for any point $(c_1,...,c_N,g_1,g_2,...)$ satisfying, say, the condition $B=1$ for $B$ a product of some subset of the $g_j$, flipping the sign of any of the coordinates $g_j$ appearing in the product will cause it to violate the condition, and vice versa.

Alternatively, we may visualize the sets of phase space points appearing in the sums \eqref{commops} and \eqref{anticommops} by first reassigning the outcome labels as
\begin{equation}
    c_i\rightarrow c'_i=\frac{1}{2}(1-c_i),\quad
    g_i\rightarrow g'_i=\frac{1}{2}(1-g_i),
\end{equation}
i.e., the outcome assignments are mapped as 
\begin{equation}
    1\rightarrow0,\quad-1\rightarrow1,
\end{equation}
and products of outcomes become binary sums of outcomes.
In other words, we have mapped our outcome space for each observable to $\mathbb{Z}_2$ (also denoted $\mathbb{F}_2$, the field of two elements), as is common practice in the literature (see \cite{gibbons04a}, for example).
The marginalizations \eqref{commops} and \eqref{anticommops} therefore become sums over all the joint probabilities for phase space points satisfying conditions of the form
\begin{equation}
    \label{hyperplane}
    \vec{A}\cdot\vec{c}\thinspace'+\vec{B}\cdot\vec{g}\thinspace'=1
\end{equation}
where $\vec{c}\thinspace'\equiv(c'_1,...,c'_N)$ and $\vec{g}\thinspace'\equiv(g'_1,g'_2,...)$, and $\vec{A},\vec{B}$ are vectors in $(\mathbb{Z}_2)^N,(\mathbb{Z}_2)^{|G|}$, respectively.
We can now see that \eqref{hyperplane} is the equation for a hyperplane in the phase space $(\mathbb{Z}_2)^{N+|G|}$, so we may think of the marginal probabilities for outcomes of individual measurements as sums of the joint probabilities over such hyperplanes \cite{gibbons04a}.

We wish to demonstrate that for any state there exists a joint probability distribution $P$ that reproduces as marginals the correct probabilities for the outcomes of any observable in $\mathcal{S}$.
Since each ontic state carries with it an outcome assignment to each observable in $\mathcal{S}$, it is enough to show that for any state, $P$ can reproduce as marginals the correct probabilities for the outcomes of any observable in $\mathcal{R}$.
The largest subsets of $\mathcal{R}$ that may be measured simultaneously are $\{C_{i1}\}\cup G$ for any $i=1,2,...,N$: let us refer to the probabilities for the joint outcomes for these as
\begin{equation}
    \label{subspaceprobs}
    P^{(\{C_{i1}\}\cup G)}(c_i,g_1,g_2,...),
\end{equation}
where the $c_i,g_1,g_2,...=\pm1$ label the joint outcomes.
Each $C_{i1}$ commutes with all operators in $G$, so for a given state we can directly determine (via Born's rule, using the appropriate projectors, or by actual measurements if the state is physical) the probabilities \eqref{subspaceprobs} (separately for each $i$).
$P$ will thus correctly reproduce as marginals the probabilities for the observables in $\mathcal{R}$ if and only if it correctly reproduces as marginals the joint outcome probabilities \eqref{subspaceprobs}, since these correspond to the largest simultaneously measurable subsets of $\mathcal{R}$.

Thus, we wish $P$ to satisfy
\begin{equation}
\begin{split}
    &P^{(\{C_{i1}\}\cup G)}(c_i,g_1,g_2,...)\\
    &=\sum_{\substack{c_j=\pm1,\\\forall j\neq i}}P(c_1,...,c_i,...,c_N,g_1,g_2,...)\label{subspacemarginals}
\end{split}
\end{equation}
for each $i=1,2,...,N$.
Since the $C_{i1}$ pairwise anticommute, the probabilities $P^{(\{C_{i1}\}\cup G)}$ and $P^{(\{C_{j1}\}\cup G)}$ for any $i\neq j$ cannot be correlated.
More specifically, the product $C_{i1}B_iC_{j1}B_j$ of the operators $C_{i1}B_i$ and $C_{j1}B_j$ (for $i\neq j$ and $B_i,B_j$ any products of operators in $G$) is not jointly measurable with $C_{i1}B_i$ and $C_{j1}B_j$, since none of the three commute with each other, and in general the product need not even be in $\overline{\mathcal{S}}$.
In any case, it has no bearing on the joint probabilities for $C_{i1}B_i$ and $C_{j1}B_j$; all that is required of the joint probability distribution $P$ is that it correctly reproduce the expressions \eqref{subspacemarginals}, which the assignment \eqref{jointprobassign} does.
Therefore, we may take the probability for the joint outcome $P(c_1,...,c_N,g_1,g_2,...)$ to be proportional to the product of the probabilities $P^{(\{C_{i1}\}\cup G)}(c_i,g_1,g_2,...)$ over all $i=1,2,...,N$.
Including the appropriate normalization gives an expression for the joint probabilities:
\begin{equation}
\label{jointprobassign}
    \frac{P(c_1,...,c_N,g_1,g_2,...)}{P^{(G)}(g_1,g_2,...)}
    =\prod_{i=1}^N\frac{P^{(\{C_{i1}\}\cup G)}(c_i,g_1,g_2,...)}{P^{(G)}(g_1,g_2,...)},
\end{equation}
where $P^{(G)}(g_1,g_2,...)$ denotes the probability of obtaining the joint outcome $(g_1,g_2,...)$ for the operators $G$ (which we can also obtain via Born's rule).

As we noted above, given any quantum state we can directly evaluate each of the probabilities on the right-hand side, since each is the probability for a joint outcome of one of the commuting sets of observables $\{C_{i1}\}\cup G$ (or just $G$).
Therefore, for any quantum state, \eqref{jointprobassign} gives a joint probability distribution $P(c_1,...,c_N,g_1,g_2,...)$ for the phase space points $(c_1,...,c_N,g_1,g_2,...)$, that reproduces as marginals all probabilities associated to the state, for observables in $\mathcal{S}$.

This discussion does not provide specific methods for constructing quasi-quantized models alternative to that presented in the main text.
What we have shown, however, is that epistemic states over the ontic states described in \cref{phasespace} may in principle describe the measurable properties of $\mathcal{S}$ for any state.
Thus, the door is open for broader sets of epistemic states than those allowed in our model as described in \cref{probdists}.
For our purposes, however, the set of epistemic states allowed in our model is sufficient.

\section{Noncontextual sub-Hamiltonians}
\label[appendix]{nonconsubhams}

For the HeH$^+$ Hamiltonian in \cite{peruzzo14a}, we performed a brute-force search over all noncontextual sub-Hamiltonians to find the one that gave the best approximation to the full Hamiltonian's ground state energy.

For the other three Hamiltonians in \cref{approximationtable}, as described in \cref{approx_sect} we used a greedy heuristic to find large noncontextual sub-Hamiltonians.
This ``one-by-one" heuristic greedily selected terms from the full Hamiltonian, in decreasing order of weight, while the set remained noncontextual.
In Tables~\ref{PeruzzoHeHplus}-\ref{KandalaBeH}, below, we list the full Hamiltonians together with the best noncontextual sub-Hamiltonians we found, as well as the diagonal terms only.
We also tried greedily adding larger subsets up to subsets of size six (four, in the BeH$_2$ Hamiltonian of \cite{kandala17a}), as well as checking for noncontextual subsets with certain fixed structures (by brute-force search over generating sets $\mathcal{R}$ where the anticommuting generators $C_{i1}$ act on fixed subsets of the qubits).
We found that none of the results outperformed the one-by-one greedy approach, and those that performed best simply reproduced the same noncontextual sub-Hamiltonian as the one-by-one greedy approach.

We obtained the ground state energies for the full Hamiltonians and diagonal sub-Hamiltonians by evaluating them directly using the OpenFermion software package \cite{openfermion}.
For the noncontextual sub-Hamiltonians, we evaluated the ground state energies using our quasi-quantized model as described in \cref{objfn_sec}, and optimizing by a brute-force search over the parameter space $E$.
We then checked the resulting ground state energies against those computed by OpenFermion, and found agreement to machine precision in all cases.

\begin{table*}[ht]
\begin{tabular}{|p{6in}|}
  \hline
  \vspace{-0.05in}
  \textbf{Diagonal terms:} \{'II': -1.46658,  'IZ': -0.39863, 'ZI': -0.39863, 'ZZ': 0.089735\}
  \vspace{0.05in}
  \\
  \hline
  \vspace{-0.05in}
  \textbf{Additional terms in noncontextual sub-Hamiltonian:} \{'XX': 0.099524\}
  \vspace{0.05in}
  \\
  \hline
  \vspace{-0.05in}
  \textbf{Additional terms in full Hamiltonian:} \{'IX': -0.087145, 'XI': -0.087145, 'XZ': 0.087145, 'ZX': 0.087145\}
  \vspace{-0.1in}
  \\
  \hline
  \vspace{-0.05in}
  \textbf{Minimal energy parameter setting for $\mathcal{R}$, in noncontextual sub-Hamiltonian:}\\
  $q_1=\langle ZZ\rangle\mapsto+1,\quad r_1=\langle XX\rangle\mapsto-0.1238712791070418,\quad r_2=\langle IZ\rangle\mapsto0.9922982949760547$
  \vspace{0.05in}
  \\
  \hline
\end{tabular}
\caption{HeH$^+$ Hamiltonian in Peruzzo \emph{et al.}, 2014 \cite{peruzzo14a}. Terms are displayed in the format \{Pauli operator:coefficient,...\}, i.e., as a Python dict mapping Pauli operators that appear in the Hamiltonian to their coefficients. Coefficients are given in Hartree.\label{PeruzzoHeHplus}}
\end{table*}

\begin{table*}[ht]
\begin{tabular}{|p{6in}|}
  \hline
  \vspace{-0.05in}
  \textbf{Diagonal terms:} \{'III':-6.823060333, 'ZII':-0.1110098029, 'IZI':-0.5370907285, 'IIZ':-0.3127149146, 'ZZI':0.383637914, 'ZIZ':0.2581256772, 'IZZ':0.2523178271\}
  \vspace{0.05in}
  \\
  \hline
  \vspace{-0.05in}
  \textbf{Additional terms in noncontextual sub-Hamiltonian:} \{'XXI':0.06593809513, 'YYI':-0.06593809513\}
  \vspace{0.05in}
  \\
  \hline
  \vspace{-0.05in}
  \textbf{Additional terms in full Hamiltonian:} \{'XIX':0.0121680127, 'YIY':0.0121680127, 'IXX':0.01764480014, 'IYY':0.01764480014\}
  \vspace{0.05in}
  \\
  \hline
  \vspace{-0.05in}
  \textbf{Minimal energy parameter setting for $\mathcal{R}$, in noncontextual sub-Hamiltonian:}\\
  $q_1=\langle IIZ\rangle\mapsto+1,\quad q_2=\langle ZZI\rangle\mapsto-1,$\\
  $r_1=\langle XXI\rangle\mapsto-0.000000129227,\quad r_2=\langle ZII\rangle\mapsto-0.999999999999$
  \vspace{0.05in}
  \\
  \hline
\end{tabular}
\caption{LiH Hamiltonian in Hempel \emph{et al.}, 2018 \cite{hempel18a}. Terms are displayed in the format \{Pauli operator:coefficient,...\}, i.e., as a Python dict mapping Pauli operators that appear in the Hamiltonian to their coefficients. Coefficients are given in Hartree.\label{HempelLiH}}
\end{table*}

\begin{table*}[ht]
\begin{tabular}{|p{6in}|}
  \hline
  \vspace{-0.05in}
  \textbf{Diagonal terms:} \{'ZIII':-0.096022, 'IZII':0.364746, 'IIZI':0.096022, 'IIIZ':-0.364746, 'ZZII':-0.206128, 'ZIZI':-0.145438, 'ZIIZ':0.110811, 'IZZI':0.110811, 'IZIZ':-0.095216, 'IIZZ':-0.206128, 'ZZZI':-0.056040, 'ZZIZ':0.063673, 'ZIZZ':0.056040, 'IZZZ':-0.063673, 'ZZZZ':0.080334\}
  \vspace{0.05in}
  \\
  \hline
  \vspace{-0.05in}
  \textbf{Additional terms in noncontextual sub-Hamiltonian:} \{'YYZI': 0.039155, 'XXZI': -0.039155, 'YYII': 0.02964, 'XXII': -0.02964, 'YYIZ': -0.02428, 'XXIZ': 0.02428, 'YYZZ': 0.002895, 'XXZZ': -0.002895\}

  \vspace{0.05in}
  \\
  \hline
  \vspace{-0.05in}
  \textbf{Additional terms in full Hamiltonian:} \{'XZII':-0.012585, 'XIII':0.012585, 'IIXZ':0.012585, 'XZIZ':0.007265, 'XIIZ':-0.007265, 'XZZI':-0.011962, 'XIZI':0.011962, 'XZZZ':-0.000247, 'XIZZ':0.000247,'IIZZ':0.012585, 'IIXI':0.012585, 'XZXZ':-0.002667, 'XZXI':-0.002667, 'XIXZ':0.002667, 'XIXI':0.002667, 'IZXZ':0.007265, 'IZXI':0.007265, 'IXII':0.002792, 'IIXX':-0.029640, 'IIIX':0.002792, 'XIXX':-0.008195, 'XIIX':-0.001271, 'XXXI':-0.008195, 'XXXX':0.028926, 'XXIX':0.007499, 'IXXI':-0.001271, 'IXXX':0.007499, 'IXIX':0.009327, 'IIYY':0.029640, 'YYYY':0.028926, 'ZXII':0.002792, 'IIZX':-0.002792, 'ZIZX':-0.016781, 'ZIIX':0.016781, 'ZXZI':-0.016781, 'IXZI':-0.016781, 'ZXZX':-0.009327, 'ZXIX':0.009327, 'IXZX':-0.009327, 'ZIXZ':-0.011962, 'ZIXI':-0.011962, 'ZZXZ':0.000247, 'ZZXI':0.000247, 'ZIXX':0.039155, 'ZZXX':-0.002895, 'ZZIX':-0.009769, 'IZXX':-0.024280, 'IZIX':-0.008025, 'ZIYY':-0.039155, 'ZZYY':0.002895, 'IZYY':0.024280, 'XZXX':0.008195, 'XZIX':0.001271, 'XZYY':-0.008195, 'XIYY':0.008195, 'XZZX':-0.001271, 'XIZX':0.001271, 'IZZX':0.008025, 'IXZZ':-0.009769, 'IXIZ':0.008025, 'XXXZ':-0.008195, 'IXXZ':-0.001271, 'YYXZ':0.008195, 'YYXI':0.008195, 'XXYY':-0.028926, 'IXYY':-0.007499, 'YYXX':-0.028926, 'YYIX':-0.007499, 'XXZX':-0.007499, 'YYZX':0.007499, 'ZZZX':0.009769, 'ZXXZ':-0.001271, 'ZXXI':-0.001271, 'ZXIZ':0.008025, 'ZXXX':0.007499, 'ZXYY':-0.007499, 'ZXZZ':-0.009769\}
  \vspace{0.05in}
  \\
  \hline
  \vspace{-0.05in}
  \textbf{Minimal energy parameter setting for $\mathcal{R}$, in noncontextual sub-Hamiltonian:}\\
  $q_1=\langle IIIZ\rangle\mapsto+1,\quad q_2=\langle ZZII\rangle\mapsto+1,\quad q_3=\langle IIZI\rangle\mapsto-1,$\\
  $r_1=\langle YYZI\rangle\mapsto-0.2192200361485217,\quad r_2=\langle IZII\rangle\mapsto-0.9756754459096738$
  \vspace{0.05in}
  \\
  \hline
\end{tabular}
\caption{LiH Hamiltonian in Kandala \emph{et al.}, 2017 \cite{kandala17a}. Terms are displayed in the format \{Pauli operator:coefficient,...\}, i.e., as a Python dict mapping Pauli operators that appear in the Hamiltonian to their coefficients. Coefficients are given in Hartree.\label{KandalaLiH}}
\end{table*}

\begin{table*}[ht]
\begin{tabular}{|p{6in}|}
  \hline
  \vspace{-0.05in}
  \textbf{Diagonal terms:} \{'ZIIIII':-0.143021, 'ZZIIII':0.104962, 'IZZIII':0.038195, 'IIZIII':-0.325651, 'IIIZII':-0.143021, 'IIIZZI':0.104962, 'IIIIZZ':0.038195, 'IIIIIZ':-0.325651, 'IZIIII':0.172191, 'ZZZIII':0.174763, 'ZIZIII':0.136055, 'ZIIZII':0.116134, 'ZIIZZI':0.094064, 'ZIIIZZ':0.099152, 'ZIIIIZ':0.123367, 'ZZIZII':0.094064, 'ZZIZZI':0.098003, 'ZZIIZZ':0.102525, 'ZZIIIZ':0.097795, 'IZZZII':0.099152, 'IZZZZI':0.102525, 'IZZIZZ':0.112045, 'IZZIIZ':0.105708, 'IIZZII':0.123367, 'IIZZZI':0.097795, 'IIZIZZ':0.105708, 'IIZIIZ':0.133557, 'IIIIZI':0.172191, 'IIIZZZ':0.174763, 'IIIZIZ':0.136055\}
  \vspace{0.05in}
  \\
  \hline
  \vspace{-0.05in}
  \textbf{Additional terms in noncontextual sub-Hamiltonian:} \{'IIZIZX': 0.010064, 'IIZIIX': -0.010064, 'ZIIIIX': -0.009922, 'ZIIIZX': 0.009922, 'IZZIIX': 0.007952, 'IZZIZX': -0.007952, 'ZZIIIX': 0.007016, 'ZZIIZX': -0.007016, 'IIIZZX': -0.002246, 'IIIZIX': 0.002246\}

  \vspace{0.05in}
  \\
  \hline
  \vspace{-0.05in}
  \textbf{Additional terms in full Hamiltonian:} \{'XZIIII':0.059110, 'XIIIII':-0.059110, 'IZXIII':0.161019, 'IIXIII':-0.161019, 'IIIXZI':0.059110, 'IIIXII':-0.059110, 'IIIIZX':0.161019, 'IIIIIX':-0.161019, 'XIXIII':-0.038098, 'XZXIII':-0.003300, 'XZIXZI':0.013745, 'XZIXII':-0.013745, 'XIIXZI':-0.013745, 'XIIXII':0.013745, 'XZIIZX':0.011986, 'XZIIIX':-0.011986, 'XIIIZX':-0.011986, 'XIIIIX':0.011986, 'IZXXZI':0.011986, 'IZXXII':-0.011986, 'IIXXZI':-0.011986, 'IIXXII':0.011986, 'IZXIZX':0.013836, 'IZXIIX':-0.013836, 'IIXIZX':-0.013836, 'IIXIIX':0.013836, 'IIIXIX':-0.038098, 'IIIXZX':-0.003300, 'ZZXIII':-0.002246, 'ZIXIII':0.002246, 'ZIIXZI':0.014815, 'ZIIXII':-0.014815, 'ZZIXZI':-0.002038, 'ZZIXII':0.002038, 'XIZIII':-0.006154, 'XZZIII':0.006154, 'XZIZII':0.014815, 'XIIZII':-0.014815, 'XZIZZI':-0.002038, 'XIIZZI':0.002038, 'XZIIZZ':0.001124, 'XIIIZZ':-0.001124, 'XZIIIZ':0.017678, 'XIIIIZ':-0.017678, 'YIYIII':-0.041398, 'YYIXXZ':0.011583, 'YYIIXI':-0.011094, 'IYYXXZ':0.010336, 'IYYIXI':-0.005725, 'IIIXIZ':-0.006154, 'XXZXXZ':0.011583, 'XXZIXI':-0.011094, 'IXIXXZ':-0.011094, 'IXIIXI':0.026631, 'IIZXII':-0.017678, 'XXZYYI':0.011583, 'XXZIYY':0.010336, 'IXIYYI':-0.011094, 'IXIIYY':-0.005725, 'IIIYIY':-0.041398, 'YYIYYI':0.011583, 'YYIIYY':0.010336, 'IYYYYI':0.010336, 'IYYIYY':0.010600, 'XXZXXX':0.024909, 'IXIXXX':-0.031035, 'XXZYXY':0.024909, 'IXIYXY':-0.031035, 'YYIXXX':0.024909, 'IYYXXX':0.021494, 'YYIYXY':0.024909, 'IYYYXY':0.021494, 'XXZZXZ':0.011094, 'IXIZXZ':-0.026631, 'YYIZXZ':0.011094, 'IYYZXZ':0.005725, 'XXZZXX':0.010336, 'IXIZXX':-0.005725, 'YYIZXX':0.010336, 'IYYZXX':0.010600, 'XXXXXZ':0.024909, 'XXXIXI':-0.031035, 'IIXIIZ':-0.010064, 'XXXYYI':0.024909, 'XXXIYY':0.021494, 'YXYXXZ':0.024909, 'YXYIXI':-0.031035, 'YXYYYI':0.024909, 'YXYIYY':0.021494, 'XXXXXX':0.063207, 'XXXYXY':0.063207, 'YXYXXX':0.063207, 'YXYYXY':0.063207, 'XXXZXZ':0.031035, 'IIXZII':-0.009922, 'YXYZXZ':0.031035, 'XXXZXX':0.021494, 'YXYZXX':0.021494, 'ZXZXXZ':0.011094, 'ZXZIXI':-0.026631, 'ZXZYYI':0.011094, 'ZXZIYY':0.005725, 'ZXZXXX':0.031035, 'ZXZYXY':0.031035, 'ZXZZXZ':0.026631, 'ZXZZXX':0.005725, 'ZXXXXZ':0.010336, 'ZXXIXI':-0.005725, 'ZXXYYI':0.010336, 'ZXXIYY':0.010600, 'ZXXXXX':0.021494, 'ZXXYXY':0.021494, 'ZXXZXZ':0.005725, 'ZXXZXX':0.010600, 'IZZXZI':0.001124, 'IZZXII':-0.001124, 'IIZXZI':0.017678, 'IZXZII':0.009922, 'IZXZZI':-0.007016, 'IIXZZI':0.007016, 'IZXIZZ':-0.007952, 'IIXIZZ':0.007952, 'IZXIIZ':0.010064, 'IIIXZZ':0.006154\}
  \vspace{0.05in}
  \\
  \hline
  \vspace{-0.05in}
  \textbf{Minimal energy parameter setting for $\mathcal{R}$, in noncontextual sub-Hamiltonian:}\\
  $q_1=\langle ZZZIII\rangle\mapsto-1,\quad q_2=\langle IIIZII\rangle\mapsto+1,\quad q_3=\langle IZIIII\rangle\mapsto-1,$\\
  $q_4=\langle IIIIZI\rangle\mapsto-1,\quad q_5=\langle ZIIIII\rangle\mapsto+1,$\\
  $r_1=\langle IIIIZX\rangle\mapsto-0.7522001251805058,\quad r_2=\langle IIIIIZ\rangle\mapsto0.6589347248995392$
  \vspace{0.05in}
  \\
  \hline
\end{tabular}
\caption{BeH$_2$ Hamiltonian in Kandala \emph{et al.}, 2017 \cite{kandala17a}. Terms are displayed in the format \{Pauli operator:coefficient,...\}, i.e., as a Python dict mapping Pauli operators that appear in the Hamiltonian to their coefficients. Coefficients are given in Hartree.\label{KandalaBeH}}
\end{table*}

\end{document}